\documentclass[aps, prb, preprint, showpacs, amssymb, amsmath, superscriptaddress]{revtex4-1}

\usepackage{mathrsfs}
\usepackage{bm}
\usepackage{times}
\usepackage{graphicx}
\usepackage[colorlinks,citecolor=blue,linkcolor=blue]{hyperref}
\usepackage{array}
\usepackage{float}
\usepackage{colortbl}
\usepackage{multirow}
\usepackage{tabularx}

\begin{document}

\title{Type-I and type-II Nodal Lines Coexistence in the Antiferromagnetic monolayer CrAs$_{2}$}

\author{Busheng Wang}
\affiliation{State Key Laboratory of Metastable Materials Science and Technology \& Key Laboratory for Microstructural Material Physics of Hebei Province, School of Science, Yanshan University, Qinhuangdao, 066004, China}
\author{Heng Gao}
\affiliation{Department of Physics, Shanghai University, 99 Shangda Road, Shanghai, 200444, China}
\author{Qing Lu}
\affiliation{Institute of Atomic and Molecular Physics, College of Physical
Science and Technology, Sichuan University, Chengdu 610065, China}
\author{WenHui Xie}
\email{whxie@phy.ecnu.edu.cn}\affiliation{Department of Physics, East China Normal University, Shanghai 200062, China}
\author{Yanfeng Ge}
\affiliation{State Key Laboratory of Metastable Materials Science and Technology \& Key Laboratory for Microstructural Material Physics of Hebei Province, School of Science, Yanshan University, Qinhuangdao, 066004, China}
\author{Yong-Hong Zhao}
\affiliation{College of Physics and Electronic Engineering, Center for Computational Sciences, Sichuan Normal University, Chengdu, 610068, China}
\author{Kaicheng Zhang}
\affiliation{Department of Physics, Bohai University, Jinzhou 121000, China}
\author{Yong Liu}
\email{yongliu@ysu.edu.cn, ycliu@ysu.edu.cn}\affiliation{State Key Laboratory of Metastable Materials Science and Technology \& Key Laboratory for Microstructural Material Physics of Hebei Province, School of Science, Yanshan University, Qinhuangdao, 066004, China}

\date{\today}

\begin{abstract}

Topological nodal line semimetals, hosting one-dimensional Fermi lines with symmetry protection, has become a hot topic in topological quantum matter. Due to the breaking of time reversal symmetry in magnetic system, nodal lines require protection by additional symmetries. Here, we report the discovery of antiferromagnetic type-I and type-II nodal lines coexist in the monolayer CrAs$_{2}$ based on a systematic first-principles calculation. Remarkably, the type-I nodal line in CrAs$_{2}$ form a concentric loop centered around the $\Gamma$ point is filling-enforced by nonsymmorphic analogue symmetry and robust against spin-orbital coupling. The type-II nodal lines, a kind of open nodal lines appear around the Fermi level, are protected by the mirror symmetry in the absence of spin-orbital coupling. The antiferromagnetic monolayer CrAs$_{2}$ proposed here may provide a platform for the correlation between magnetism and exotic topological phases.

\end{abstract}
 \maketitle

\section*{Introduction}
The discovery of Dirac cones in graphene\cite{gra1,gra2,gra3,gra4,gra5,quan1,quan2} have inspired a continuous research of Dirac semimetals (DSMs)\cite{QHE1,QHE2,kl,magnet,DSM1,DSM2,DSM4,DSM6,DSM9,DSM11}. The massless Dirac fermions in DSMs play the key role in diverse quantum phenomena, such as the novel quantum Hall effect\cite{QHE1,QHE2}, Klein tunneling\cite{kl}, and giant linear magnetoresistance\cite{magnet} and so on. DSMs are characterized by the point nodes or nodal lines (NLs) where the conduction and valence bands cross in the Brillouin zone(BZ). The nodal line semimetals (NLSMs) present line band crossing with no dispersion along the NL direction and with linear dispersion in the perpendicular direction under certain crystalline symmetries\cite{DNL1,DNL2,DNL4,DNL5,DNL6,DNL7,DNL8,DNL9,PbTaSe2,DNL10,DNL11,DNL12,DNL13,DNL14,DNL15,DNL17,DNL18,CuS2}. Depending on the degree of tilting, NLSMs can be classified into type-I and type-II\cite{tII-1,tII-2}. Previous studies on NLSMs have focused on time reversal symmetry $\mathcal{T}$ invariant systems with and without spin-orbit coupling (SOC). Symmetry breaking may lead NLSMs into different exotic topological states such as topological insulators\cite{PbTaSe2,MX,AFM3} and nodal point semimetals\cite{PD1,PD2}. Generally, the formation of magnetic order is accompanied with $\mathcal{T}$ symmetry breaking and sometimes followed by a decrease in crystalline symmetry. Therefore, it is challenging to find NLSMs in a magnetic materials.

Recently, a few theoretical and experimental works started exploring the nodal features in antiferromagnetic (AFM) systems\cite{AFM3,DNL16,Jwang,Jwang2,Young2017,AFM1,AFe2As2}. The AFM phase of CuMnAs\cite{AFM3} was predicted as a three-dimensional (3D) NLSM where the NLs are created by the band-inversion transition, and can be gapped in the presence of SOC. More recent theoretical works\cite{DNL16,Jwang2} report a distinctive class of 3D NLSM in AFM systems: essential NLSM. The NLs are filling-enforced by the combination of Kramers theorem and nonsymmorphic analogue symmetry $\tilde{\mathcal{T}}$ = ${\left \{\mathcal{T}| \textbf{\emph{t}}\right \}}$, where $\textbf{\emph{t}}$ is a fraction of the lattice vector. Furthermore, unlike the NLs in band-inversion NLSMs, the NLs in essential NLSMs are robust against SOC due to their gaplessness. Based on the above, it is worth asking whether NLs can be obtained in two-dimensional AFM materials without $\mathcal{T}$ symmetry. To the best of our knowledge, the real two-dimensional (2D) AFM NLSMs have not been reported. Considering the great success in the field of graphene, it is expected that searching for 2D AFM NLSMs will most likely lead to the discovery of numerous noteworthy physical phenomena and novel topological states.

In this work, we performed electronic structure calculations and demonstrated that monolayer CrAs$_{2}$, an overlooked 2D material, hosts AFM type-I and type-II NLs in the absence of SOC. It is worth noting that the essential type-I NL is proposed in 2D AFM materials for the first time. This NL is filling-enforced by nonsymmorphic analogue symmetry $\tilde{\mathcal{T}}$ and cannot be gapped even in the presence of SOC. The type-II NLs, formed by band-inversion between the $d$$_{x^{2}-y^{2},xy}$ and $d$$_{xz,yz}$ orbitals of the transition metal Cr, are protected by the mirror symmetry $M_z$ in the absence of SOC. Moreover, the topological nontriviality was verified by the flat edge states. In addition, the feasibility of exfoliation from CrAs$_{2}$ layered bulk phases was computationally predicted based on small cleavage energies. The phonon calculations and \emph{ab} \emph{initio} molecular dynamics simulations suggest that freestanding monolayer CrAs$_{2}$ is dynamically and thermally stable. The AFM monolayer CrAs$_{2}$ proposed here may provide a platform for the realization of relation between magnetism and exotic topological phases.

\section*{COMPUTATIONAL DETAILS}
First-principles calculations were performed using the projected-augmented-wave (PAW) method\cite{PAW} as implemented in the Vienna \emph{ab} \emph{initio} simulation package (VASP)\cite{vasp1,vasp2}. The exchange-correlation energy was treated using Perdew-Burke-Ernzerhof (PBE)\cite{PBE} generalized gradient approximation, and the optB88-vdW\cite{vdw1,vdw2} dispersion correction was applied to account for the long-range van der Waals interactions. A vacuum region of 15 ${\mbox{\AA}}$ was set to avoid the interactions between the adjacent atomic layers and its period image could be neglected. A kinetic cutoff energy of 550 eV was used for the expansion of the wave-function and Monkhorst-Pack k meshes with a grid spacing of $2\;\pi \times 0.03$ {\mbox{\AA}$^{-1}$ to ensure that the enthalpy converges to better than 1 $\times$  10$^{-5}$ eV per cell. The calculation of the phonon spectra was performed using the density functional perturbation theory\cite{linres1,linres2,linres3} with VASP and PHONOPY codes\cite{phonopy1,phonopy2}. Moreover, \emph{ab} \emph{initio} molecular dynamics (AIMD) simulations with canonical ensemble using the Nos$\acute{e}$ heat bath scheme were performed to evaluate the thermal stability. To study the topological edge states, a first-principles tight-binding model Hamiltonian was constructed by projecting onto the Wannier orbitals\cite{wannier1,wannier2,wannier3} with the VASP2WANNIER90 interface\cite{vaspwannier}. The Cr $d$ and As $p$ orbitals were used to build the maximally localized Wannier functions, and then we calculated the edge states using the iterative Green's function method as implemented in the WannierTools package\cite{wanniertools}.

\section*{RESULTS AND DISCUSSION}
\subsection{Structure, cleavage and stability }

Monolayer CrAs$_{2}$ crystallizes in a 1H structure with a $D_{3h}^{1}$ point group symmetry. Each monolayer consists of a hexagonal plane of Cr atoms sandwiched between two hexagonal planes of As atoms, as shown in Fig.~\ref{fig:latt}(a). The Cr atoms are trigonal prismatically coordinated and bonded by the six nearest-neighboring As atoms. This particular stacking does not preserve the space inversion symmetry $\mathcal{P}$. However, with respect to the Cr atomic plane, the lattice is reflection-symmetric under the mirror operation $\emph{M}_{z}$. Considering that the transition metal atoms can induce magnetism in materials\cite{Mag2,Mag3}, the energies of three different configurations (nonmagnetic (NM), ferromagnetic (FM), and AFM) was compared to determine the preferred magnetic ground state of the monolayer CrAs$_{2}$, as presented in Figure S1 of the Supplemental Information. It's found that the optimized AFM state is the most energetically stable at the [001]-direction magnetization, the magnetic moment is 2.0 $\mu$B per Cr atom. Fig.~\ref{fig:latt}(a) shows that the spin-polarization of the monolayer CrAs$_{2}$ can be characterized by the spin polarized charge density. All the magnetic moments are clearly localized around the Cr atoms, and the AFM configuration dominates the magnetic ground state of the monolayer CrAs$_{2}$. The optimized lattice parameters for the AFM unit cell are $\emph{a}$ = $\emph{b}$ = 7.44 $\mbox{\AA}$, $\emph{d}_{As-As}$ = 2.46 $\mbox{\AA}$ and $\emph{d}_{Cr-As}$ = 2.48 $\mbox{\AA}$, respectively. In this AFM system, both the spatial symetry $C_{3}$ and time-reversal symmetry $\mathcal{T}$ are broken. However, a nonsymmorphic analogue symmetry $\tilde{\mathcal{T}}$ = ${\left \{\mathcal{T}| \textbf{\emph{t}}\right \}}$ is induced by the AFM order, where $\textbf{\emph{t}}$ = (1/2, 0, 0) is a vector connecting the two spin-polarized sublattices. This nonsymmorphic analogue symmetry $\tilde{\mathcal{T}}$ and the reflection symmetry $\emph{M}_{z}$ of the lattice provide protection for the topological NLs, as discussed below.

\begin{figure*}[htp!]
\centerline{\includegraphics[width=0.9\textwidth]{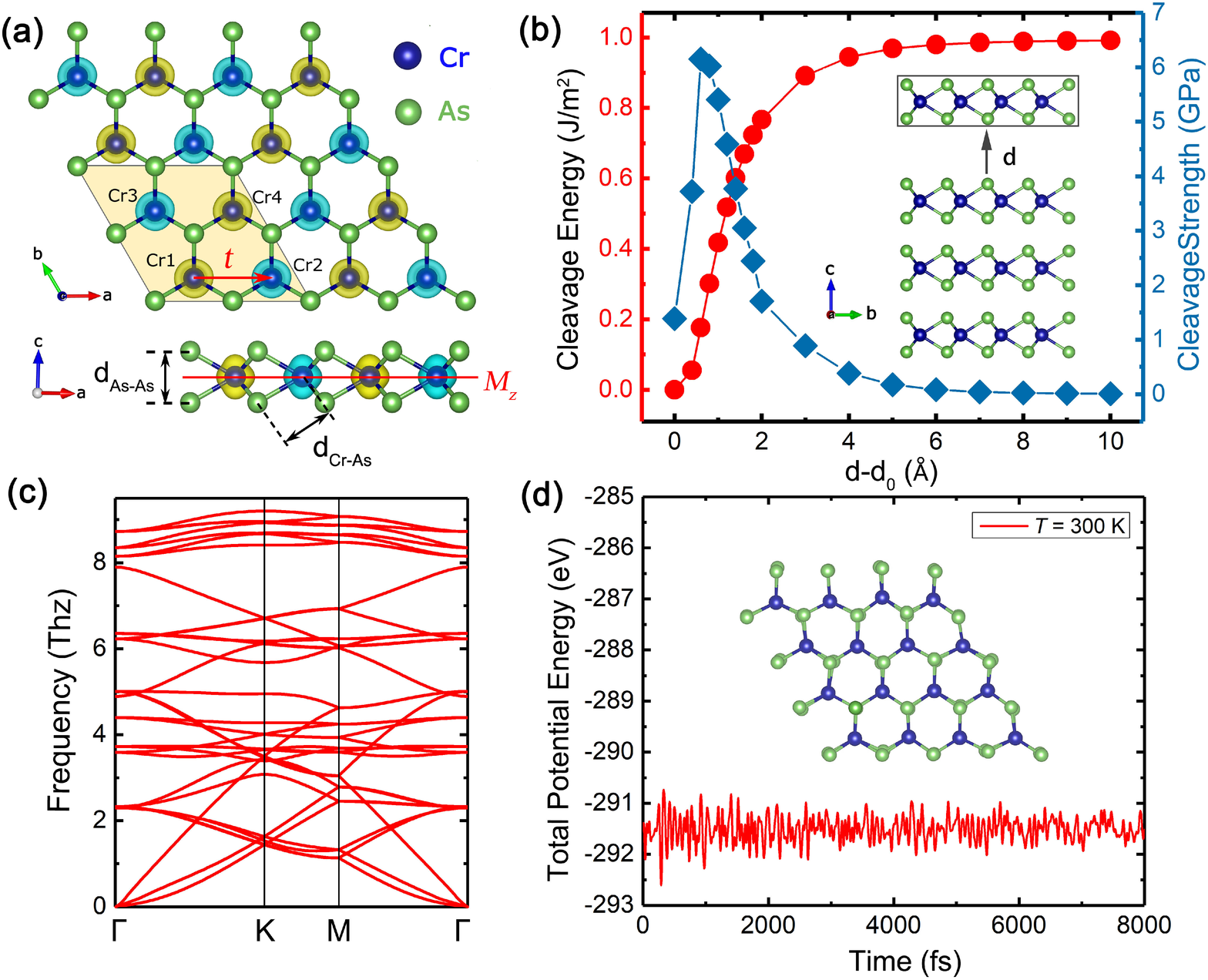}}
\caption{(Color online) (a) Crystal structure of the monolayer CrAs$_{2}$ (top view) the unit cell of the AFM configuration is highlighted in orange. There are four Cr atoms (blue spheres) and eight As atoms (green spheres) in each unit cell. The yellow and cyan isosurfaces correspond to the spin-up and spin-down components, respectively, in the spatial spin density distributions. $\textbf{\emph{t}}$ = (1/2, 0, 0) is the half translation along the (100) direction. (b) The cleavage energy and strength calculated using the optB88-vdW functional as a function of the separation d between two fractured parts. d$_{0}$ represents the equilibrium interlayer distance; (c) Phonon dispersion of monolayer CrAs$_{2}$ calculated using the PBE functional; (d) Total energy fluctuations with respect to molecular dynamics simulation step observed for the monolayer CrAs$_{2}$ at 300 K.}
\label{fig:latt}
\end{figure*}

Then we focused on the possibility of CrAs$_{2}$ layers exfoliated from the bulk and the stability of the freestanding monolayer. \emph{Ab} \emph{initio} calculations predict that the bulk crystal CrAs$_{2}$ with a $P\bar{6}m2$ (No. 187) space group is a metastable compound, as depicted in Fig. S2 of the Supplemental Material. The bulk CrAs$_{2}$ consists of weakly van der Waals bonded monolayers 1H-CrAs$_{2}$ with AA stacking along the $\emph{c}$ axis. The quasi-2D nature of the bulk CrAs$_{2}$ enables the creation of a stable monolayer 1H-CrAs$_{2}$ by micromechanical cleavage and liquid exfoliation\cite{excl1,excl2}.

To estimate the feasibility of obtaining monolayers, the cleavage energies were calculated as a function of the separation between two fractured parts using the optB88-vdW functional, as shown in Fig.~\ref{fig:latt}(b). A large distance between two layers representing a fracture in the bulk, inset of Fig.~\ref{fig:latt}(b) was introduced to simulate the exfoliation procedure\cite{cl1,cl2}. The energy relative to the uncleaved equilibrium state increases with increasing separation $d$, and gradually converges to the cleavage energy. The calculated cleavage energy for CrAs$_{2}$ is 0.99 J/m$^{2}$, which locates between those of ReSe$_{2}$ (1.10 J/m$^{2}$)\cite{ReSe2}and graphite (0.37 J/m$^{2}$)\cite{graphite1,graphite2}. Considering that the exfoliation of monolayers ReSe$_{2}$ and graphene in the experiments\cite{ReSe22,graphite2}, the same is expected for CrAs$_{2}$. Furthermore, the absence of any imaginary frequencies of phonon spectrum in the entire BZ confirms the dynamical stability of the monolayer, and we also performed AIMD simulations for the monolayer CrAs$_{2}$ at room temperature, as shown in Fig.~\ref{fig:latt}(d). The atomic configuration of the monolayer CrAs$_{2}$ including the trigonal prismatical CrAs$_{6}$ and honeycomb networks remain unchanged. Clearly, the predicted low cleavage energy, dynamic and thermal stability of the monolayer suggest that the freestanding monolayer can be obtained experimentally even at room temperature.

\begin{figure}[htp!]
\centerline{\includegraphics[width=0.9\textwidth]{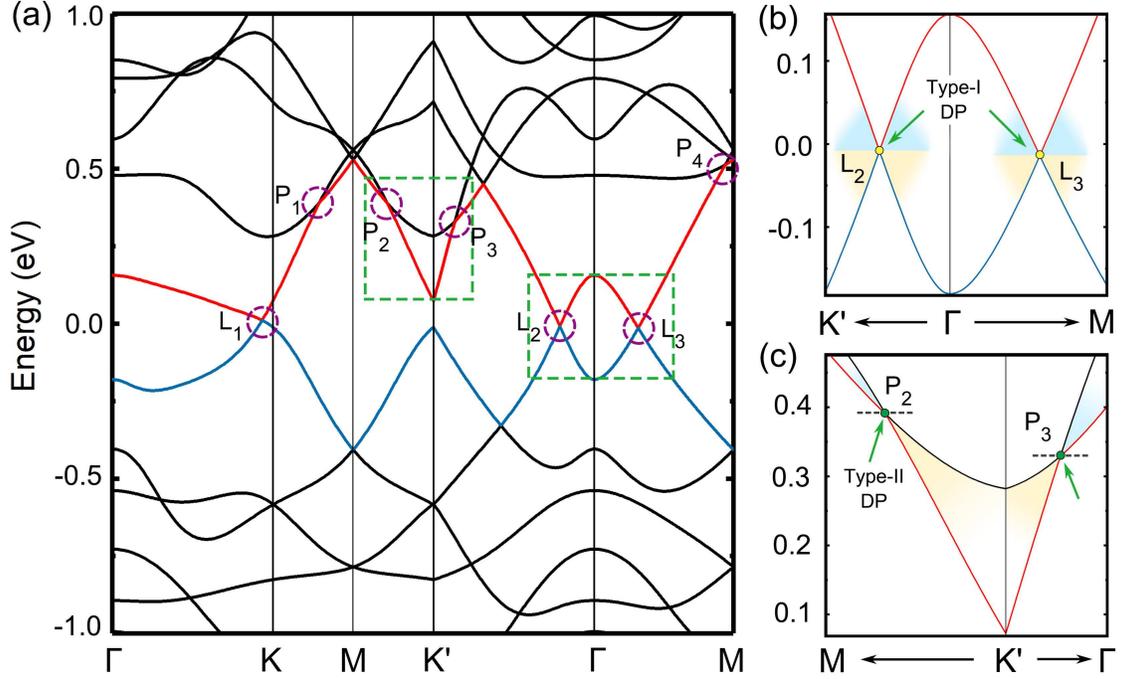}}
\caption{(Color online) (a) The electronic band structures of the monolayer CrAs$_{2}$ without SOC. Dirac cones are denoted by L$_{1-3}$ and P$_{1-4}$, which correspond to type-I and type-II Dirac points, respectively. An zoom-in of the band structure within the green boxes is shown in (b) and (c). The valence band maximum and conduction band minimum are represented as blue and red lines. The Fermi level is set to zero.}
\label{fig:2d}
\end{figure}

\subsection{Electronic properties.}

Having identified the magnetic ground state and having assessed the stability of the monolayer CrAs$_{2}$, then we investigated the detailed electronic properties. First, the spin-polarized electronic band structure of the monolayer in the absence of SOC is discussed. As Fig.~\ref{fig:2d}(a) shows, we found that the material displays a semimetal band structure: the valence and conduction bands meet in the vicinity of the Fermi level and show three band crossings, labeled as L$_{i}$ ($i$ = 1, 2, 3). Four additional band crossings are observed above the Fermi level within the energy range of 0.6 eV, P$_{i}$ ($i$ = 1, 2, 3, 4). Considering the band dispersion slope, these linear crossing points L$_{i}$ and P$_{i}$ are Type-I and type-II Dirac points, respectively.

\begin{figure*}[htp!]
\centerline{\includegraphics[width=1.0\textwidth]{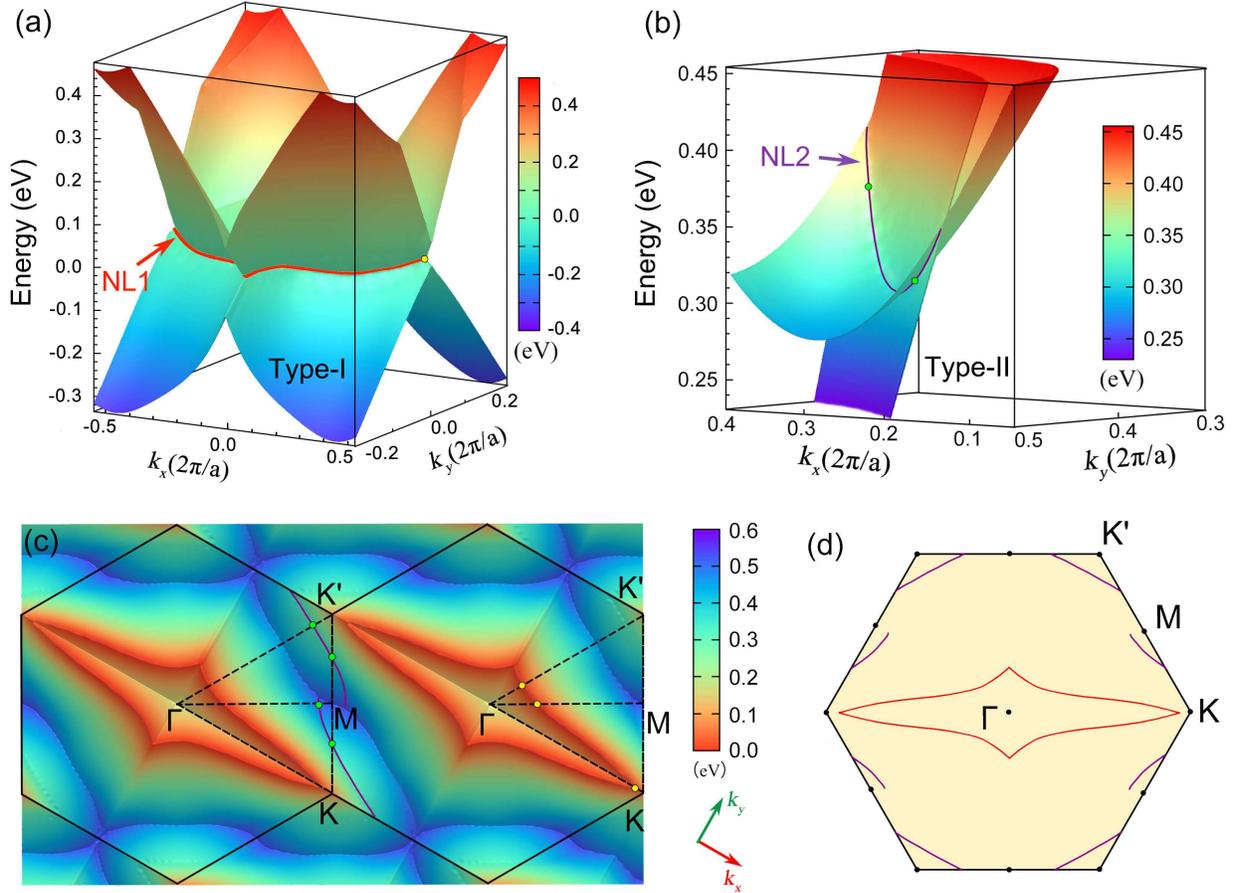}}
\caption{(Color online) (a) 3D energy band of closed NL (type-I) around L$_{i}$ ($i$ = 1, 2, 3); (b) 3D energy band of open NL (type-II) around P$_{i}$ ($i$ = 2, 3); (c) Projection of CBM in the BZ, and the Dirac cones at the 2D band (Fig.~\ref{fig:2d}(a)) are denoted by the yellow (type-I) and green (type-II) circles. The direction of the reciprocal space is indicated by the arrow magenta, the color represents the energy of the 3D bands; (d) Scheme showing the shape of the NLs in the whole BZ. The Type-I and type-II NLs are denoted by red and purple lines, respectively. The Fermi level is set to zero.}
\label{fig:td}
\end{figure*}

To further understand the Dirac features in the whole BZ, the 3D band dispersions for the monolayer CrAs$_{2}$ are depicted in Fig.~\ref{fig:td}. The valence band maximum (VBM) and conduction band minimum (CBM) meet in the vicinity of the Fermi level with opposite slopes and show a closed type-I NL, as shown in Fig.~\ref{fig:td}(a). The closed NL does not lie flat on the Fermi level but has a finite dispersion in the region near the this level. Fig.~\ref{fig:td}(b) clearly shows a type-II NL around the P$_{i}$ ($i$ = 2, 3) band crossings (the two bands are tilted in such a way that they share the same sign in their slope
along one transverse direction). For further analysis of the distribution of NLs, the 3D band of the CBM in the reciprocal space is shown in Fig.~\ref{fig:td}(c). An exotic electronic structure with type-I NLs (red lines) is observed, and these NLs form a continuous loop around the $\varGamma$ points. In addition, open type-II NLs are located around ${K}$ points, as depicted in purple along ${K}$-${K}'$ line. The Dirac points mentioned in the 2D band (Fig.~\ref{fig:2d}), locating at the Type-I and type-II NLs, are represented by the yellow and green dots in  Fig.~\ref{fig:td}. Furthermore, Fig.~\ref{fig:td}(d) schematically shows the coexistence of Type-I and type-II NLs in the whole BZ. Such coexistence around Fermi level suggests that the ultrahigh mobility in the monolayer CrAs$_{2}$ is not only confined to Fermi level, but also above it within the energy range of 0.6 eV like to the commonly phenomenon in Dirac materials\cite{MF3}.

\begin{figure*}[htp!]
\centerline{\includegraphics[width=1.0\textwidth]{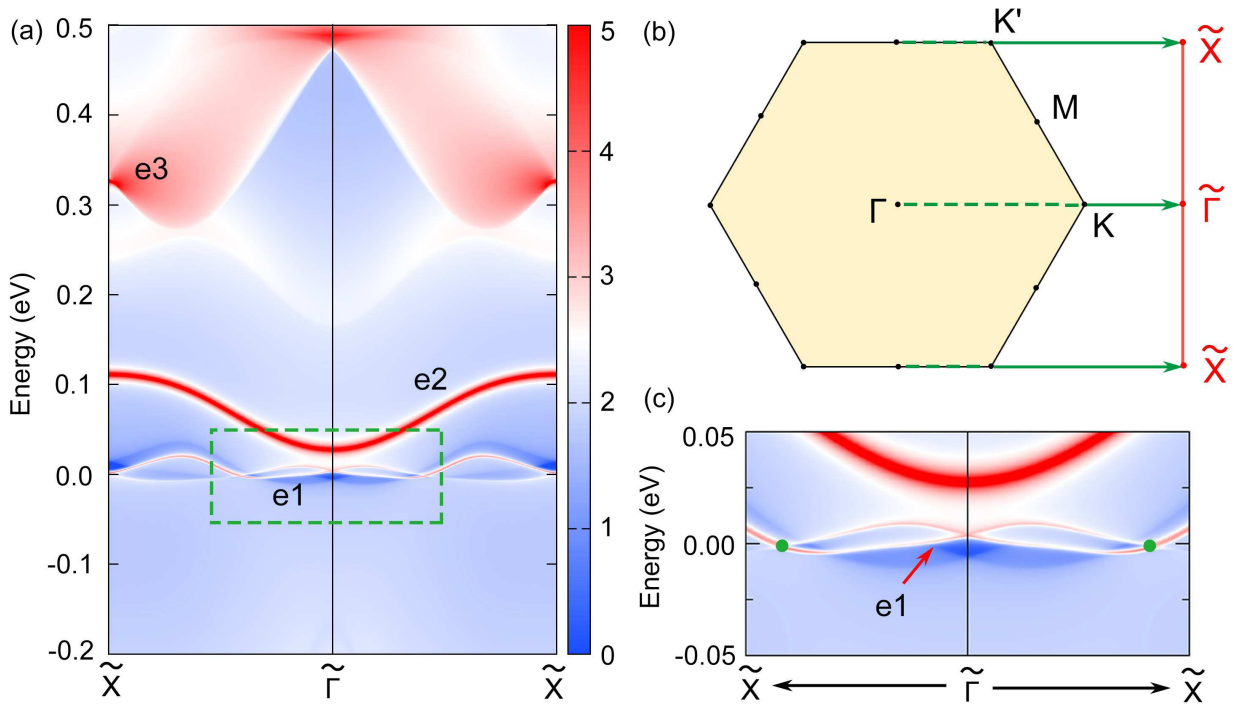}}
\caption{(Color online) (a) Projected spectrum on the (010) surface of the monolayer CrAs$_{2}$, the edge states are denoted as e1 - e3, and the correspondence between them and the semimetal states are e1 $\rightarrow$ L$_{i}$, e3 $\rightarrow$ P$_{i}$;  e2 is a normal edge state. (b) 2D BZ of the monolayer CrAs$_{2}$ and its projected 1D BZ. (c) Enlarged spectrum of edge state e1. The green dots indicate the location of the protected band crossing points on the type-I nodal loop. Color indicates the charge density for each state at the surface layer.}
\label{fig:ss}
\end{figure*}

Because the topological stable NLSM state have nontrivial edge states\cite{SS1,SS2}, we calculated the band spectrum of the (010) edge, as shown in Fig.~\ref{fig:ss}. According to the energy range of semimetal states, as shown in Fig.~\ref{fig:2d}, the correspondence between these edge and semimetal states are: e1 $\rightarrow$ L$_{i}$, e3 $\rightarrow$ P$_{i}$. The edge state e1 is split because of the inversion symmetry broken,and the double degeneracy is lifted for the edge bands of the type-I nodal loop, similar to Dirac NL materials\cite{SSsplit1,SSsplit2}. Furthermore, after detailed analysis of the edge states with different symmetry, we found that e2 corresponds to a normal edge states derived from the symmetry broken of the ribbon edge.

\subsection{The stability of the Dirac nodal lines}

To elucidate the physical origin and symmetry related properties of the type-I and type-II NLs, we analyzed the orbital composition of the states near the NLs, then artificially broke certain spatial symmetries to test the robustness of the NLs against SOC. Fig. S3 in the Supplemental Material depicts the orbitally resolved band structures of the monolayer CrAs$_{2}$. The type-I NLs in the monolayer CrAs$_{2}$ predominantly originate from the in-plane Cr-$d$$_{x^{2}-y^{2},xy}$ orbitals, the type-II NLs are due to the band-inversion between $d$$_{x^{2}-y^{2},xy}$ and $d$$_{xz,yz}$, as shown in Fig.~\ref{fig:SDNL}(b). However, $d$-orbital Dirac materials generally have a stronger spin exchange interaction and a larger SOC, which may result in novel physical properties\cite{CuS2,PbTaSe2}. Fig.~\ref{fig:SDNL}(c) shows the band structure with SOC for the monolayer CrAs$_{2}$. Small gaps (18, 23 and 25 meV along $\varGamma$-${K}$, ${K}'$-$\varGamma$ and $\varGamma$-${M}$, respectively) are observed around the type-II NL2 under SOC. Interestingly, we found band crossings in the vicinity of the Fermi level without opening of energy gaps, thus the NL1 gaplessness is robust against SOC.

\begin{figure*}[htp!]
\centerline{\includegraphics[width=1.0\textwidth]{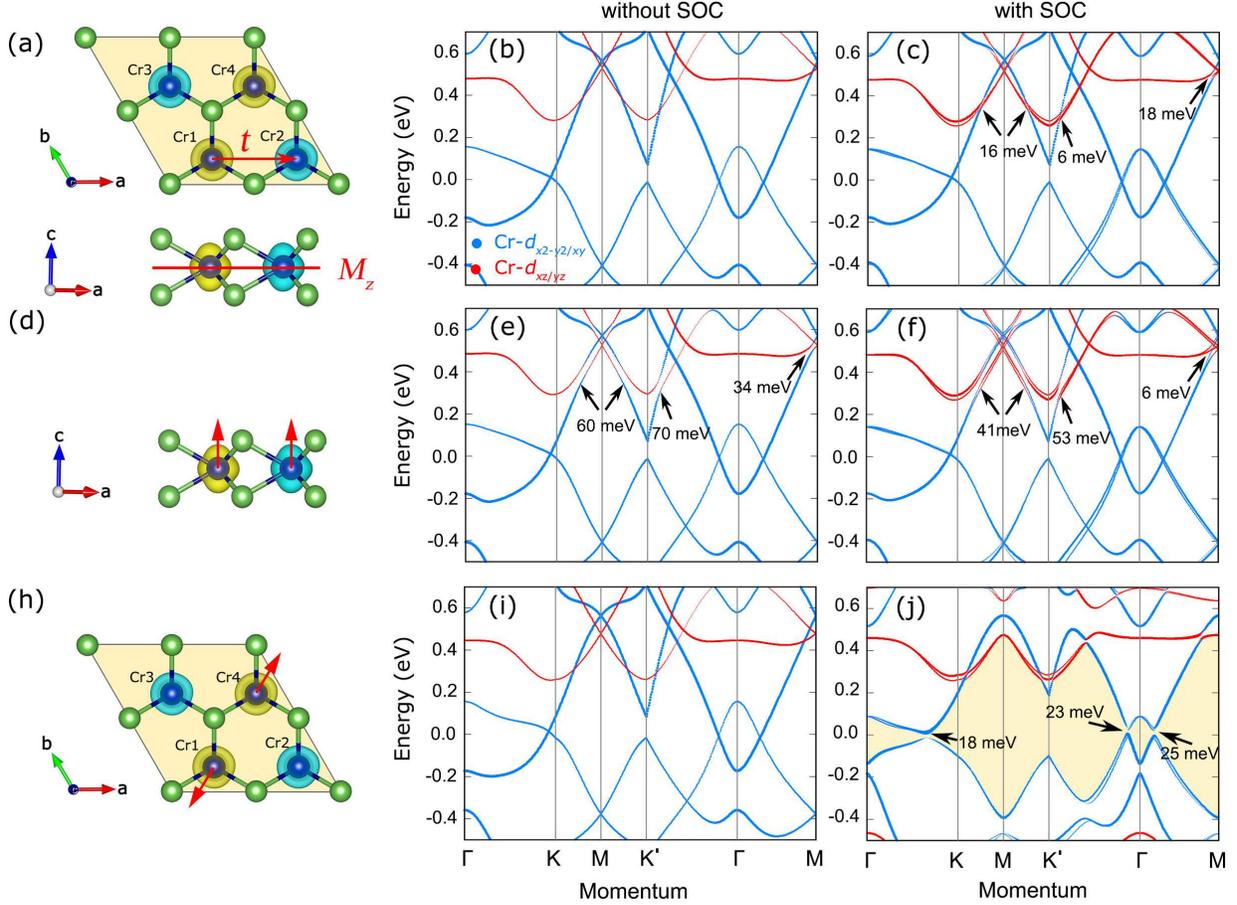}}
\caption{(Color online) (a) Top and side view of the lattice structure for the AFM monolayer CrAs$_{2}$ with both nonsymmorphic analogue $\tilde{\mathcal{T}}$ = ${\left \{\mathcal{T}|(1/2, 0, 0)\right \}}$ and mirror $M_z$ symmetry. (d) Side view of the lattice structure after breaking the $M_z$ symmetry but maintaining the $\tilde{\mathcal{T}}$ symmetry. Red arrows indicate the shift (0.1 $\mbox{\AA}$) of the Cr atoms along the $z$-axis to break $M_z$. (h) Top view of the lattice structure after breaking the $\tilde{\mathcal{T}}$ symmetry but maintaining the $M_z$ symmetry. Red arrows indicate the shift (0.1 $\mbox{\AA}$) of Cr1 and Cr4 atoms along (-1, -1, 0) and (1, 1, 0) direction to break $\tilde{\mathcal{T}}$. The band structured calculated without (b, e, i) and wit (c, f, j) SOC for the configurations in (a), (d) and (h), respectively. The orbital components. are shown in either red or blue. The Fermi level is set to zero.}
\label{fig:SDNL}
\end{figure*}

To demonstrate whether the two NLs in the monolayer CrAs$_{2}$ are symmetry protected, we performed two tests to evaluate symmetry related properties of the NLs. First, we artificially shifted the positions of all the Cr atoms along the $z$-axis to break the mirror symmetry $M_z$, while maintaining the nonsymmorphic analogue symmetry $\tilde{\mathcal{T}}$ = ${\left \{\mathcal{T}|((1/2), 0, 0)\right \}}$, as shown in Fig.~\ref{fig:SDNL}(d). The corresponding electronic structure without SOC clearly shows that the NL2 is fully gapped (Fig.~\ref{fig:SDNL}(e)). Contrastingly, the gapless NL1 could be observed even in the presence of SOC (Fig.~\ref{fig:SDNL}(f)). The evidence shows that type-II NL2 are symmetry protected by mirror symmetry in the absence of SOC.

In the second test, we artificially shifted the positions of the Cr1 and Cr4 atoms along the (-1, -1, 0) and (1, 1, 0) direction, respectively, to break $\tilde{\mathcal{T}}$ while maintaining the $M_z$ symmetry, as shown in Fig.~\ref{fig:SDNL}(h). The band structure with SOC demonstrates that NL1 is fully gapped. The values for the gap along $\varGamma$-${K}$, ${K}'$-$\varGamma$ and $\varGamma$-${M}$ are 18, 23 and 25 meV, respectively. The symmetry mechanism for essential NLs has been discussed extensively for AFM systems\cite{DNL16,Jwang2}. It should be noted that, such NL is essential, and cannot be gapped without lowering the magnetic space group symmetries. These results confirm that the essential NL1 is robust against SOC and filling-enforced by the nonsymmorphic analogue symmetry (the combination of AFM time-reversal symmetry and fractional translation).

\section*{Conclusion}
We found that type-I and type-II NLs coexist in the low energy electronic state of an AFM monolayer CrAs$_{2}$ without SOC. The essential type-I NL is proposed in AFM 2D materials for the first time. This NL is filling-enforced by nonsymmorphic analogue symmetry, and cannot be gapped even in the presence of SOC. The type-II NLs, formed by the band-inversion between the $d$$_{x^{2}-y^{2},xy}$ and $d$$_{xz,yz}$ orbitals of the Cr atoms, are protected by the mirror symmetry in the absence of SOC. Furthermore, the topological nontriviality was verified by the flat edge states on the boundary. In addition, the feasibility of exfoliation from CrAs$_{2}$ layered bulk phases was computationally predicted based on the small cleavage energies. The phonon spectrum and AIMD simulation suggest that freestanding monolayer CrAs$_{2}$ is dynamically and thermally stable. Thus the AFM monolayer CrAs$_{2}$ proposed here may provide a platform for the correlation between magnetism and exotic topological phases. We believe that the results of this work are suitable for experimental verification and potential application.

\begin{acknowledgments}
This work was supported by the NSFC under Grants  No. 11747054, 11504027, 11104191 and 51572086. Y. Liu acknowledges support by the Natural Science Foundation of Hebei Province (No. A2015203021). We thank Dr. R. Yu for useful discussions.

\end{acknowledgments}

%\newpage
%\section*{References}


\begin{thebibliography}{99}
\bibitem{gra1}
A. H. C. Neto, N. M. R. Peres, K. S. Novoselov, and A. K. Geim,
Rev. Mod. Phys. \textbf{81}, 109 (2009).

\bibitem{gra2}
A. K. Geim and K. S. Novoselov,
Nat. Mater. \textbf{6}, 183 (2007).

\bibitem{gra3}
G. W. Semenoff,
Phys. Rev. Lett. \textbf{53}, 2449 (1984).

\bibitem{gra4}
K. S. Novoselov, A. K. Geim, S. V. Morozov, D. Jiang, M. I.
Katsnelson, I. V. Grigorieva, S. V. Dubonos, and A. A.
Firsov,
Nature \textbf{438}, 197 (2005).

\bibitem{gra5}
M. I. Katsnelson, K. S. Novoselov, and A. K. Geim,
Nat. Phys. \textbf{2}, 620 (2006).

\bibitem{quan1}
C. L. Kane and E. J. Mele,
Phys. Rev. Lett. \textbf{95}, 226801 (2005).

\bibitem{quan2}
C. L. Kane and E. J. Mele,
Phys. Rev. Lett. \textbf{95}, 146802 (2005).

\bibitem{QHE1}
K. S. Novoselov, A. K. Geim, S. V. Morozov, D. Jiang, Y.
Zhang, S. V. Dubonos, I. V. Grigorieva, and A. A. Firsov,
Science \textbf{306}, 666 (2004).

\bibitem{QHE2}
Y. Zhang, Y. W. Tan, H. L. Stormer, and P. Kim,
Nature \textbf{438}, 201 (2005).

\bibitem{kl}
C. W. J. Beenakker,
Rev. Mod. Phys. \textbf{80}, 1337 (2008).

\bibitem{magnet}
T. Liang, Q. Gibson, M. N. Ali, M. Liu,
R. J. Cava, and N. P. Ong,
Nat. Mater. \textbf{14}, 280 (2015).

\bibitem{DSM1}
S. M. Young, S. Zaheer, J. C. Y. Teo, C. L. Kane, E. J. Mele, and A. M. Rappe,
Phys. Rev. Lett. \textbf{108}, 140405 (2012).

\bibitem{DSM2}
Z. Wang, Y. Sun, X.-Q. Chen, C. Franchini, G. Xu, H. Weng, X. Dai, and Z. Fang,
Phys. Rev. B \textbf{85}, 195320 (2012).

\bibitem{DSM4}
J. A. Steinberg, S. M. Young, S. Zaheer, C. L. Kane, E. J. Mele, and A. M. Rappe,
Phys. Rev. Lett. \textbf{112}, 036403 (2014).

\bibitem{DSM6}
Z. K. Liu, B. Zhou, Y. Zhang, Z. J. Wang, H. M. Weng, D.
Prabhakaran, S.-K. Mo, Z. X. Shen, Z. Fang, X. Dai, Z. Hussain, and Y. L. Chen,
Science \textbf{343}, 864 (2014).

\bibitem{DSM9}
B. J. Wieder, Y. Kim, A. M. Rappe, and C. L. Kane,
Phys. Rev. Lett. \textbf{116}, 186402 (2016).

\bibitem{DSM11}
H. Watanabe, H. C. Po, M. P. Zaletel, and A. Vishwanath,
Phys. Rev. Lett. \textbf{117}, 096404 (2016).


\bibitem{DNL1}
A. A. Burkov, M. D. Hook, and L. Balents,
Phys. Rev. B \textbf{84}, 235126 (2011).

\bibitem{DNL2}
C.-K. Chiu and A. P. Schnyder,
Phys. Rev. B \textbf{90}, 205136 (2014).

\bibitem{DNL4}
H. Weng, Y. Liang, Q. Xu, R. Yu, Z. Fang, X. Dai, and Y. Kawazoe,
Phys. Rev. B \textbf{92}, 045108 (2015).

\bibitem{DNL5}
Y. Kim, B. J. Wieder, C. L. Kane, and A. M. Rappe,
Phys. Rev. Lett. \textbf{115}, 045108 (2015).

\bibitem{DNL6}
R. Yu, H. Weng, Z. Fang, X. Dai, and X. Hu,
Phys. Rev. Lett. \textbf{115}, 036807 (2015).

\bibitem{DNL7}
C. Fang, Y. Chen, H.-Y. Kee, and L. Fu,
Phys. Rev. B \textbf{92}, 081201 (2015).

\bibitem{DNL8}
K. Mullen, B. Uchoa, and D. T. Glatzhofer,
Phys. Rev. Lett. \textbf{115}, 026403 (2015).

\bibitem{DNL9}
R. Li, H. Ma, X. Cheng, S. Wang, D. Li, Z. Zhang, Y. Li, and X.-Q. Chen,
Phys. Rev. Lett. \textbf{117}, 096401 (2016).

\bibitem{PbTaSe2}
G. Bian, T.-R. Chang, R. Sankar, S.-Y. Xu, H. Zheng, T. Neupert, C.-K. Chiu, S.-M. Huang, G. Chang, I. Belopolski, D. S. Sanchez, M. Neupane, N. Alidoust, C. Liu, B. Wang, C.-C. Lee, H.-T. Jeng, C. Zhang, Z. Yuan, S. Jia, A. Bansil, F. Chou, H. Lin, and M. Z. Hasan,
Nat. Commun. \textbf{7}, 10556 (2016).

\bibitem{DNL10}
J. Hu, Z. Tang, J. Liu, X. Liu, Y. Zhu, D. Graf, K. Myhro, S. Tran, C. N. Lau, J. Wei, and Z. Mao,
Phys. Rev. Lett. \textbf{117}, 016602 (2016).

\bibitem{DNL11}
L. M. Schoop, M. N. Ali, C. Straer, A. Topp, A. Varykhalov, D. Marchenko, V. Duppel, S. S. P. Parkin, B. V. Lotsch, and C. R. Ast,
Nat. Commun. \textbf{7}, 11696 (2016).

\bibitem{DNL12}
Y. Wu, L.-L. Wang, E. Mun, D. D. Johnson, D. Mou, L. Huang, Y. Lee, S. L. Budo, P. C. Canfield, and A. Kaminski,
Nat. Phys. \textbf{12}, 667 (2016).

\bibitem{DNL13}
M. Hirayama, R. Okugawa, T. Miyake, and S. Murakami, Nat. Commun. \textbf{8}, 14022 (2017).

\bibitem{DNL14}
Q. Xu, R. Yu, Z. Fang, X. Dai, and H. Weng, Phys. Rev. B \textbf{95}, 045136 (2017).

\bibitem{DNL15}
Q.-F. Liang, J. Zhou, R. Yu, Z. Wang, and H. Weng, Phys. Rev. B \textbf{93}, 085427 (2016).


\bibitem{DNL17}
S. Guan, Y. Liu, Z.M. Yu, S. S. Wang, Y. Yao, and S. A. Yang, Phys. Rev. Mater. \textbf{1}, 054003 (2017).

\bibitem{DNL18}
S. Li, Y. Liu, S. S. Wang, Z. M. Yu, S. Guan, X. L. Sheng, Y. Yao, and S. A. Yang, Phys. Rev. B, \textbf{97}, 045131 (2018).

\bibitem{CuS2}
B. Feng, B. Fu, S. Kasamatsu, S. Ito, P. Cheng, C.-C. Liu, S. K. Mahatha, P. Sheverdyaeva, P. Moras, M. Arita, et al.
Nat. Commun. \textbf{8}, 1007 (2017).


\bibitem{tII-1}
S. Li, Z. M. Yu, Y. Liu, S. Guan, S.S. Wang, X. Zhang, Y. Yao, A. S. Yang,
Phys. Rev. B, \textbf{96}, 081106 (2017).

\bibitem{tII-2}
X. Zhang, L. Jin, X. Dai, and G. Liu, J. Phys. Chem. Lett. \textbf{8}, 4814-4819 (2017).


\bibitem{MX}
Y. J. Jin, R. Wang, J. Z. Zhao,  Y. P. Du, C. D. Zheng, L. Y. Gan, J. F. Liu, H. Xu, and S. Y. Tong,
Nanoscale \textbf{9}, 13112-13118. (2017).

\bibitem{AFM3}
P. Tang, Q. Zhou, G. Xu, and S.-C. Zhang, Nat. Phys. \textbf{12}, 1100 (2016).

\bibitem{PD1}
C. Fang, H. Weng, X. Dai, and Z. Fang, Chin. Phys. B \textbf{25}, 117106 (2016).

\bibitem{PD2}
J. Liu, D. Kriegner, L. Horak, D. Puggioni, C. R. Serrao, R. Chen, D. Yi, C. Frontera, V. Holy, A. Vishwanath, and J.M. Rondinelli,
Phys. Rev. B \textbf{93}, 085118 (2016).

\bibitem{AFM1}
J. Park, G. Lee, F. Wolff-Fabris, Y. Y. Koh, M. J. Eom, Y. K. Kim, M. A. Farhan, Y. J. Jo, C. Kim, J. H. Shim,
and J. S. Kim, Phys. Rev. Lett. \textbf{107}, 126402 (2011).

\bibitem{AFe2As2}
Z.-G. Chen, L. Wang, Y. Song, X. Lu, H. Luo, C. Zhang, P. Dai, Z. Yin, K. Haule, and G. Kotliar,
Phys. Rev. Lett. \textbf{119}, 096401 (2017).

\bibitem{Jwang}
J. Wang, Phys. Rev. B \textbf{95}, 115138 (2017).

\bibitem{Young2017}
S. M. Young and C. L. Kane, Phys. Rev. Lett. \textbf{118}, 186401 (2017).

\bibitem{DNL16}
T. Bzdusek, Q. Wu, A. Ruegg, M. Sigrist, and A. A. Soluyanov,
Nature (London) \textbf{538}, 75 (2016).

\bibitem{Jwang2}
J. Wang,
Phys. Rev. B \textbf{96}, 081107 (2017).

\bibitem{PAW}
P. E. Blochl, Phys. Rev. B \textbf{50}, 17953 (1994).

\bibitem{vasp1}
G. Kresse, and J. Furthmuller, Phys. Rev. B \textbf{54}, 11169 (1996).

\bibitem{vasp2}
G. Kresse, and J. Furthmuller, Comput. Mater.Sci. \textbf{6}, 15 (1996).

\bibitem{PBE}
J. P. Perdew, K.Burke, and  M. Ernzerhof,
Phys. Rev.Lett. \textbf{77}, 3865 (1996).

\bibitem{vdw1}
J. Klimese, D. R. Bowler, and A. Michaelides,
Phys. Rev. B \textbf{83}, 195131 (2011).

\bibitem{vdw2}
J. Klimese, D. R. Bowler, and A. Michaelides,
J. Phys.: Condens. Mat. \textbf{22}, 022201 (2010).

\bibitem{linres1}
S. Baroni, P. Giannozzi, and A. Testa,
Phys. Rev. Lett. \textbf{58}, 1861 (1987).

\bibitem{linres2}
S. Baroni,  S. De Gironcoli, A. Dal Corso, and P. Giannozzi,
Rev. Mod. Phys. \textbf{73}, 515 (2001).

\bibitem{linres3}
X. Gonze,
Phys. Rev. A \textbf{52}, 1096 (1995).

\bibitem{phonopy1}
A. Togo, F. Oba, and I. Tanaka,
Phys. Rev. B \textbf{78}, 134106 (2008).

\bibitem{phonopy2}
A. Togo, L. Chaput, I. Tanaka, , and G. Hug,
Phys. Rev. B \textbf{81}, 174301 (2010).

\bibitem{wannier1}
N. Marzari, and D. Vanderbilt,
Phys. Rev. B \textbf{56}, 12847 (1997).

\bibitem{wannier2}
I. Souza, N. Marzari, and D. Vanderbilt,
Phys. Rev. B \textbf{65}, 035109 (2001).

\bibitem{wannier3}
A. A. Mosto, J. R. Yates, Y. S. Lee, I. Souza, D. Vanderbilt, and N. Marzari,
Comput. Phys. Commun. \textbf{178}, 685 (2008).

\bibitem{vaspwannier}
C. Franchini, R. Kovik, M. Marsman, S. S. Murthy, J. He, C. Ederer, and G. Kresse,
J. Phys.: Condens. Matter \textbf{24}, 235602 (2012)

\bibitem{wanniertools}
Q. Wu, S. Zhang, H.-F. Song, M. Troyer and A. A. Soluyanov,
Comput. Phys. Commun. \textbf{224}, 405 (2018).


\bibitem{Mag2}
Y. Matsumoto, M. Murakami, T. Shono, T. Hasegawa, T. Fukumura, M. Kawasaki, P. Ahmet, T. Chikyow, S. Y. Koshihara, and H. Koinuma,
Science, \textbf{291}, 5505  (2001).

\bibitem{Mag3}
B. Wang, Q. Lu, Y. Ge, K. Zhang, W. Xie, W. M. Liu, and Y. Liu,
Phys. Rev. B \textbf{96}, 13 (2017).

\bibitem{SI}
See Supplemental Material at http://link.aps.org/supplemental/ for more detailed computational methods, as well as the bingding energy for magnetic configurations, formation enthalpies of bulk $P\bar{6}m2$ CrAs$_{2}$, orbitally resolved band structures.


\bibitem{excl1}
C. Lee, H. Yan, L. E. Brus, T. F. Heinz, J. Hone, and S. Ryu,
ACS Nano \textbf{4}, 2695 (2010).

\bibitem{excl2}
J. N. Coleman, M. Lotya, A. ONeill, S. D. Bergin, P. J. King, U. Khan, K. Young, A. Gaucher, S. De, R. J. Smith, I. V. Shvets, S. K. Arora, G. Stanton, H.-Y. Kim, K. Lee, G. T. Kim, G. S. Duesberg, T. Hallam, J. J. Boland, J. J. Wang, J. F. Donegan, J. C. Grunlan, G. Moriarty, A. Shmeliov, R. J. Nicholls, J. M. Perkins, E. M. Grieveson, K. Theuwissen, D. W. McComb, P. D. Nellist, and V. Nicolosi,
Science \textbf{331}, 568 (2011).

\bibitem{cl1}
N. I. Medvedeva, O. N. Mryasov, Y. N. Gornostyrev, D. L. Novikov, and A. J. Freeman,
Phys. Rev. B \textbf{54}, 13506 (1996).

\bibitem{cl2}
B. Sachs, T. O. Wehling, K. S. Novoselov, A. I. Lichtenstein, and M. I. Katsnelson,
Phys. Rev. B \textbf{88}, 201402 (2013).

\bibitem{ReSe2}
Y. Jiao, L. Zhou, F. Ma, G. Gao, L. Kou, J. Bell, S. Sanvito, and A. Du,
ACS Appl. Mater. Interfaces, \textbf{8}, 5385 (2016).

\bibitem{ReSe22}
E. Liu, Y. Fu, Y. Wang, Y. Feng, H. Liu, X. Wan, W. Zhou, B. Wang, L. Shao, C. Ho, Y. Huang, Z. Cao, L. Wang, A. Li, J. Zeng, F. Song, X. Wang, Y. Shi, H. Yuan, H. Hwang, Y. Cui, F. Miao, and D. Xing,
Nat. Commun. \textbf{6},6991 (2015).

\bibitem{graphite1}
R. Zacharia, H. Ulbricht, and T. Hertel,
Phys. Rev. B \textbf{69}, 155406 (2004).

\bibitem{graphite2}
W. Wen, S. Dai, X. Li, J. Yang, D. J. Srolovitz, and Q. Zheng,
Nat. Commun. \textbf{6}, (2015).

\bibitem{MF3}
Y. Jiao, F. Ma, C. Zhang, J. Bell, S. Sanvito, and A. Du,
Phys. Rev. Lett. \textbf{119}, 016403 (2017).

\bibitem{SS1}
A. A. Burkov, M. D. Hook, and Leon Balents,
Phys. Rev. B \textbf{84}, 235126 (2011).

\bibitem{SS2}
S. Ryu and Y. Hatsugai,
Phys. Rev. Lett. \textbf{89}, 077002 (2002).

\bibitem{SSsplit1}
S. Li, Y. Liu, S. S. Wang, Z. M. Yu, S. Guan, X. L. Sheng, Y. Yao, and S. A. Yang, S.A.
Phys. Rev. B \textbf{97}, 045131 (2018).


\bibitem{SSsplit2}
S. S. Wang, Y. Liu, Z. M. Yu, X. L. Sheng, and S. A. Yang,
Nat. Commun. \textbf{8}, 1844 (2017)

\bibitem{caly2}
Y. Wang, J. Lv, L. Zhu, and Y. Ma,
Comput. Phys. Commun. \textbf{183}, 2063 (2012).

\bibitem{ZB1}
H. Akinaga, and M. Mizuguchi,
J. Phys.: Condens. Mat. \textbf{16}, S5549 (2004).



\end{thebibliography}
\end{document}